\pdfoutput=1
\documentclass[sigconf, screen]{acmart}
\usepackage{listings}
\usepackage{booktabs}
\usepackage{graphicx}
\usepackage{microtype}

\usepackage{amssymb}
\usepackage{float}
\usepackage{xcolor}
\usepackage[most]{tcolorbox}
\usepackage{enumitem}
\definecolor{codered}{HTML}{C0392B}
\definecolor{codewhite}{HTML}{FAFAFA}
\definecolor{webgray}{HTML}{F8F9FA}
\definecolor{webborder}{HTML}{DADCE0}
\definecolor{checkgreen}{HTML}{137333}
\definecolor{crossred}{HTML}{C5221F}
\definecolor{tipbg}{HTML}{FEF7E0}
\definecolor{tipborder}{HTML}{FBBC04}

\newfloat{listing}{tbhp}{lol}
\floatname{listing}{Listing}

\newcommand{\CodeIn}[1]{\begin{small}\texttt{#1}\end{small}}

\title{Grounding AI Agents in Contracts: An Empirical Evaluation of Spec-Driven Test Generation}

\author{Michele Tufano, James McClure, José Cambronero, Runxiang Cheng, Sherry Y. Shi, Renyao Wei, Dorothy Chen, Franjo Ivan\v{c}i\'{c}, Livio Dalloro, Pat Rondon}
\email{\{tufanomichele, jmcclure, jcambronero, chengsam, sherryyshi, renyaow, dschen, ivancic, dalloro, rondon\}@google.com}
\affiliation{
  \institution{Google}
  \country{USA}
}

\renewcommand{\shortauthors}{M. Tufano, J. McClure, J. Cambronero, R. Cheng, S. Shi, R. Wei, D. Chen, F. Ivan\v{c}i\'{c}, L. Dalloro, and P. Rondon}

\begin{abstract}

LLM-based agents are increasingly used for coding tasks, where they have outperformed many classical approaches and scaled to repository-level tasks, such as test generation.
However, when directly prompted to generate tests, these agents can fail to reason about the code and its underlying contracts, thereby missing edge cases and behavioral boundaries that affect test quality.
To address this limitation, we propose \emph{Spec-Driven Test Generation}, where we instruct an agent to first reason about -- and explicitly document -- code pre-conditions, post-conditions, and undefined behaviors.
This intermediate semi-formal specification acts as a \textbf{cognitive scaffold} to guide subsequent test generation.
Our evaluation on production bugs from Google shows that the spec-driven agent can  deliver a 9.8 percentage points (\textit{p} = 0.0352) improvement in bug detection rate and a 2.5 percentage point (\textit{p} = 0.0034) improvement in branch coverage, compared to a traditional test generation agent baseline.
Using LLM-as-a-Judge, we further show that test suites generated by the spec-driven agent are superior to the baseline and human-authored tests in 77.8\% and 56.7\% of the cases, respectively, and demonstrated improvements on following best practices, readability, and edge-case coverage.

\end{abstract}

\ccsdesc[500]{Software and its engineering~Software verification and validation}
\ccsdesc[500]{Software and its engineering~Empirical software validation}
\ccsdesc[500]{Computing methodologies~Artificial intelligence}
\ccsdesc[500]{Software and its engineering~Specification languages}

\keywords{Spec-Driven Development, Software Testing, Design by Contract, AI Agents, Large Language Models}

\setcopyright{cc}
\setcctype{by}
\acmDOI{10.1145/3842652.3843195}
\acmYear{2026}
\copyrightyear{2026}
\acmISBN{979-8-4007-2970-6/2026/10}
\acmConference[SpecOps '26]{Proceedings of the 1st International Workshop on Specification-Driven Development Life Cycle}{October 4--9, 2026}{Oakland, CA, USA}
\acmBooktitle{Proceedings of the 1st International Workshop on Specification-Driven Development Life Cycle (SpecOps '26), October 4--9, 2026, Oakland, CA, USA}
\acmSubmissionID{splashws26specopsmain-p4-p}
\received{2026-06-29}
\received[accepted]{2026-07-16}

\begin{document}
\maketitle
\section{Introduction}
\label{sec:introduction}

Large Language Models (LLMs) and autonomous LLM-based agents are rapidly reshaping the way developers write and test code.
While agents are capable of generating many types of software artifacts, generating effective tests requires a deep understanding of the underlying codebase and its expected behavior in order to determine the correct output for a given execution state \cite{meyer2018seven,barr2014oracle}.
Nonetheless, software engineers have long developed foundations to facilitate rigorous reasoning about expected program behavior.
For example, Design by Contract (DbC)~\cite{meyer1992applying} explicitly documents behaviors in the form of pre-/post-conditions, and invariants.

Unfortunately, when agents are directly generating tests from existing code, there are no explicit contracts to anchor their exploration towards generating more effective tests.
Agents can thus miss edge cases, hallucinate logic, and generate superficial tests that fail to systematically exercise the program's state space.

To address this limitation and bridge the principles of DbC with agentic automation, we propose Spec-Driven Test Generation.
Rather than relying on human engineers to define upfront contracts for existing systems, our methodology instructs an agent to \textit{ground} its tests by first retroactively extracting and articulating a code component's underlying contract before authoring any test code.
The agent documents the pre-/post-conditions and undefined behaviors into a structured document that delineates inputs and outputs in natural language.
Because this document blends the expressiveness of natural language with structure, but does not impose a strict mathematical formalism,  we term this a \textit{semi-formal specification}.
This specification then acts as an oracle to encourage the agent to explore the state space more systematically.

To empirically evaluate whether this self-generated oracle can lead to superior test suites, we designed a comparative study evaluating a two-phase Spec-Driven Agent against an agent that directly generates test suites, which represents standard practice in agentic test generation.
On a dataset of 90 historical production bugs from Google’s Issue Tracking System, we employ a ``Greenfield Test Generation'' setup to measure improvements in structural coverage and historical bug detection rates.
Furthermore,
we evaluate the semantic completeness of the generated specs using a novel \textit{Contract Coverage} metric, which quantifies how a spec covers incorrect behavior of an evaluated bug.
To assess test rigor, we employ a pairwise LLM-as-a-Judge that uses a large LLM (Gemini 3.1 Pro) that is better capable of evaluating the outputs generated by the agents that use a small LLM (Gemini 3 Flash).

Our findings demonstrate that the Spec-Driven Agent achieves a bug detection rate of 63.2\% at \textit{k} = 5 runs, which represents a statistically significant (\textit{p} = 0.0352) 9.8 percentage points improvement over the baseline, and a highly statistically significant 2.5 percentage points improvement in branch coverage (\textit{p} = 0.0034).

Compared to the test suites generated by the baseline, a pairwise LLM-as-a-Judge rates the Spec-Driven-Agent-generated counterparts as overall superior in 77.8\% of cases and have higher quality across all evaluated categories: adherence to testing best practices (65.6\%), readability (68.9\%), and edge-case coverage (83.3\%).
Compared to the original, developer-authored test suites, the Spec-Driven-Agent-generated counterparts are rated overall superior in 56.7\% of cases, showing parity with human-level engineering rigor.

When we use our proposed Contract Coverage metric to analyze the generated specs, we find that this metric scales robustly, starting at 61.1\% at \textit{k} = 1 and reaching 78.9\% at \textit{k} = 5. This metric also represents a predictor of testing success ($p = 3.62 \times 10^{-14}$):
when the generated spec successfully captures the violated behavioral contract, the resulting test suite detects the target bug in 54.9\% of cases, compared to a mere 19.4\% detection rate when the contract is missing from the generated spec.

In summary, this paper makes the following contributions:

\begin{itemize}
\item \textbf{Spec-Driven Test Generation Framework:} We propose a novel, two-phase agentic methodology that effectively improves the agent's test generation ability
by instructing agents to explicitly extract and document behavioral contracts prior to code synthesis.

\item \textbf{Contract Coverage Metric:} We introduce a novel evaluation metric designed to measure the empirical coverage of AI-generated specifications and assess their capacity to capture violated business logic in real-world scenarios rather than claiming formal mathematical completeness.

\item \textbf{Industrial Empirical Evaluation:} We provide a rigorous comparative analysis using a dataset of 90 historical bug-fixes from Google's production systems, demonstrating that Spec-Driven Test Generation yields statistically significant improvements in both structural coverage and historical bug detection, while matching the qualitative rigor of expert, human-authored test suites.

\end{itemize}

Our findings suggest that treating semi-formal specification generation as a prerequisite step---rather than relying on direct code synthesis---is the key to unlocking agentic test generation capable of matching expert human engineering rigor.
While this paper strictly evaluates the application of these behavioral contracts for agents
, future work can explore the broader utility of these agent-generated specifications as valuable documentation and system alignment artifacts for human developers.

\section{Semi-Formal Specifications for API Behavior}
\label{sec:specification}

Specifications are widely used in software engineering and programming languages, and what exactly constitutes a specification can vary widely, depending on the ultimate goal and the intended audience (or tool).
In practice, a program specification can range from a natural language description of intended behavior and associated properties to a rigorous mathematical representation of the expected behavior.
In this work, the term ``specification'' means a semi-formal contract for an API.
We are inspired by DbC and capture pre- and post-conditions
for APIs. However, in our approach these contracts are
written in natural language and are assembled into a structured artifact (i.e., \CodeIn{.spec.md}) that will be used to reason about the underlying program during agentic test generation.

\paragraph{Formal Definition of the Specification Artifact}

To operationalize this semi-formal paradigm, we define the specification artifact for a given source code component as a structured document consisting of specification blocks for every API (public and private functions, classes, methods) extracted from the code.

For any given code unit \textit{c}, its behavioral specification $S_c$ is formalized as the tuple: $S_c = \langle D_c, Pre_c, Post_c, Sug_c \rangle$

\begin{itemize}
\item $D_c$: A natural language description summarizing the code unit's intended behavior.

\item $Pre_c$: A set of pre-conditions that define the exact state or constraints required \textit{before} execution.

\item $Post_c$: A set of post-conditions that define the guaranteed state \textit{after} execution. These encompass input validation (error handling), success values, state side-effects (e.g., database writes), and expected state variants.

\item $Sug_c$: A local subset of test suggestions explicitly tied to enforcing the conditions in $Pre_c$ or $Post_c$.
\end{itemize}

Crucially, each condition within $Pre_c$ or $Post_c$ is intended to be evaluated against the existing unit test suite and assigned one of two verification states: (i) \textit{Tested}: The condition is empirically verified by an existing test; (ii) \textit{Untested}: The condition lacks verification, directly prompting a corresponding test suggestion in $Sug_c$.

\paragraph{Generating Specifications with AI Agents}

Semi-formal specifications provide a valuable cognitive scaffold for AI agents. By forcing the LLM to systematically articulate pre-execution requirements and post-execution guarantees, the methodology imposes the structural rigidity necessary to constrain hallucinations and rigorously explore the state space. Simultaneously, expressing these constraints in natural language yields a format that is accessible for developer review, remains expressive and flexible, and preserves the option to seamlessly adapt to diverse downstream backend formalisms (e.g., Lean~\cite{de2015lean} or SMT-LIB~\cite{BarFT-SMTLIB}).

While traditional miners (e.g., Daikon or trace analysis) treat execution as ground truth---possibly formalizing buggy behavior as expected---LLM agents avoid only summarizing the existing implementation by attempting to infer true developer intent. Specifically, the agent can cross-reference local code with natural language docstrings, comments, existing tests, and caller-site API usage patterns across the repository to deduce implicit component expectations. Synthesizing a contract from these intent sources restricts semantic drift and preserves human-readability while generalizing (out-of-the-box) across diverse languages.

\paragraph{Illustrative Example}

Consider a subset of a specification generated for a hypothetical \CodeIn{CustomerDatabase.add\_customer} method. The agent analyzes the source code and existing tests to synthesize the semi-formal contract block shown in Listing~\ref{lst:illustrative_example}.

\begin{listing}[t!]
\centering
\caption{Illustrative example of a semi-formal specification. The agent extracts pre-/post-conditions and checks whether they are already tested or missing (and suggests a test).}
\label{lst:illustrative_example}
\begin{tcolorbox}[
  colback=white,
  colframe=webborder,
  coltitle=black!80,
  colbacktitle=webgray,
  title={\small\textsf{\textbf{Specification: \texttt{CustomerDatabase.add\_customer}}}},
  boxrule=0.6pt,
  arc=3pt,
  left=6pt, right=6pt, top=4pt, bottom=4pt
]
\small
\textbf{\normalsize \texttt{CustomerDatabase.add\_customer}} \hfill \textcolor{gray}{\textsf{\small Public Method}}\\[2pt]
\textcolor{black!75}{\textit{Adds a new customer to the backend, or raises an exception if there was an error.}}\\[6pt]
\textbf{\textsf{\small Pre-conditions}}
\begin{itemize}[leftmargin=12pt, topsep=1pt, itemsep=1pt, label=\textcolor{gray}{\small$\bullet$}]
  \item The database connection has been established \textcolor{gray}{(guaranteed by public API)}.
\end{itemize}
\vspace{2pt}
\textbf{\textsf{\small Post-conditions}}\\[2pt]
\textit{\textbf{\textsf{\footnotesize Input Validation}}}
\begin{itemize}[leftmargin=14pt, topsep=1pt, itemsep=3pt]
  \item[\textcolor{checkgreen}{\large\ding{51}}] Raises \texttt{InvalidArgumentError} if \texttt{customer.phone\_number} is invalid.\\
  \hspace*{2pt}\textcolor{gray}{\scriptsize Tested by \texttt{add\_customer\_invalid\_phone\_test}}
  \item[\textcolor{crossred}{\large\ding{55}}] Raises \texttt{InvalidArgumentError} if \texttt{customer.name} is empty.
  \begin{tcolorbox}[colback=tipbg, colframe=tipborder, boxrule=0.4pt, arc=2pt, left=4pt, right=4pt, top=2pt, bottom=2pt]
    \scriptsize \textbf{Tip:} Test that \texttt{InvalidArgumentError} is raised if the \texttt{customer.name} parameter is empty.
  \end{tcolorbox}
\end{itemize}
\textit{\textbf{\textsf{\footnotesize Success Scenarios}}}
\begin{itemize}[leftmargin=14pt, topsep=1pt, itemsep=1pt]
  \item[\textcolor{checkgreen}{\large\ding{51}}] Returns \texttt{None} if the customer was successfully added.\\
  \hspace*{2pt}\textcolor{gray}{\scriptsize Tested by \texttt{add\_customer\_success\_test}}
\end{itemize}
\end{tcolorbox}
\vspace{-12pt}
\end{listing}

In this example, the agent maps verified post-conditions to their corresponding test cases (\checkmark{}).
When it identifies a gap through code analysis—such as the implementation handling an empty name, but no test explicitly verifying this path—it flags the post-condition as untested (\textbf{X}) and automatically synthesizes a structural test suggestion.
This artifact serves as the blueprint reasoning plan for the subsequent test generation phase.

\begin{figure}[t]
\centering
\includegraphics[width=1\columnwidth]{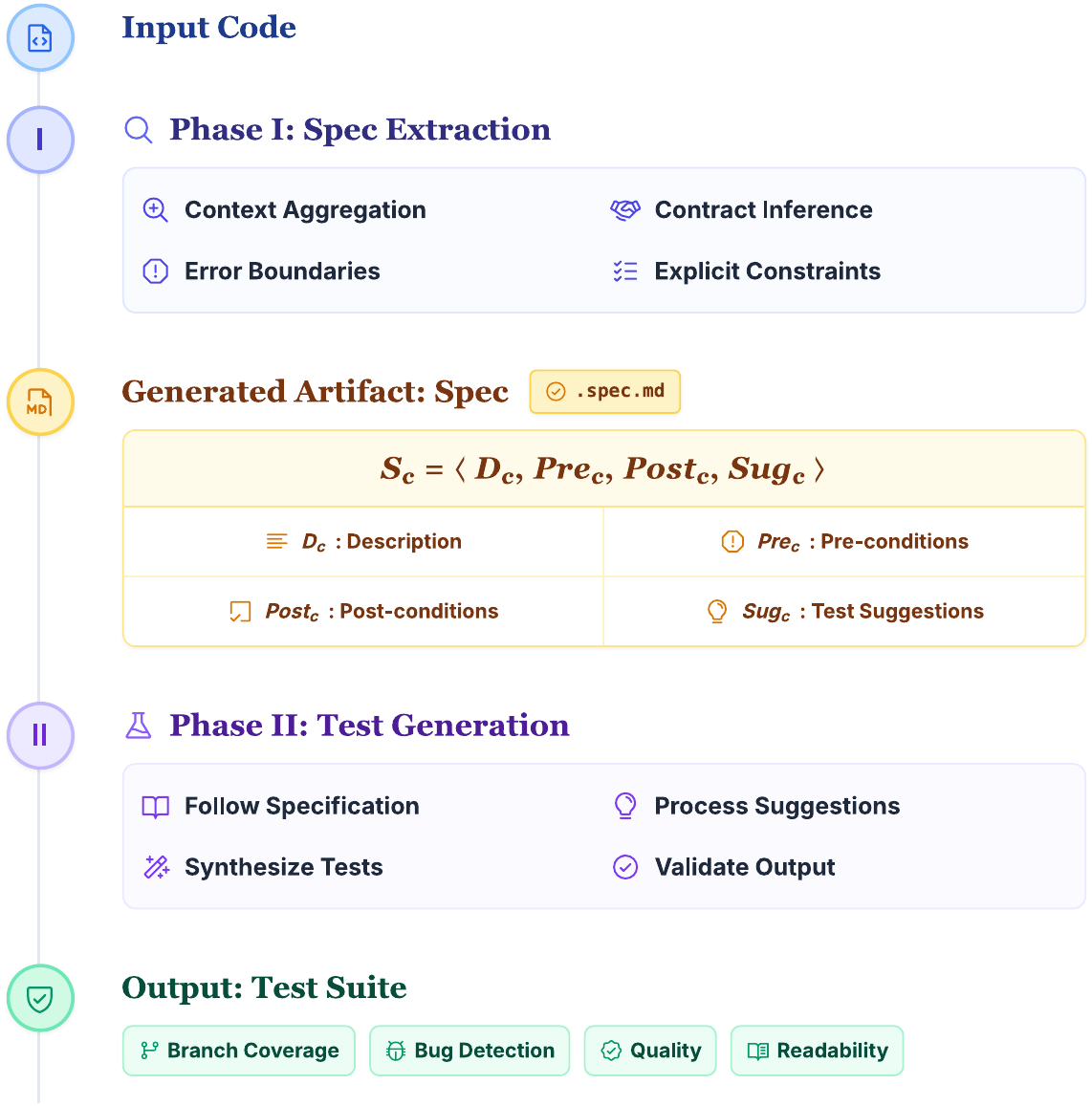}
\caption{Overview of spec-driven test generation.}
\label{fig:approach}
\end{figure}

\section{Proposed Approach}
\label{sec:design}

We now discuss our two-phase, agentic framework to generate specifications of the form described in \S\ref{sec:specification} then generate tests.
Figure~\ref{fig:approach} provides an overview of the framework.

\subsection{Phase I: Specification Extraction}
\label{sec:design:spec}

Given a target source file, optionally an existing unit test suite, and access to standard codebase navigation and editing tools, the agent's objective in the first phase is to produce a specification document containing a tuple $S_c$ for every unit of code $c$ within the component (note: we do not programmatically enforce coverage but encourage it via the agent's instructions). This phase proceeds through four steps:

\begin{enumerate}
\item \textbf{Context Aggregation:} The agent reads and analyzes the code under test—alongside any available natural language requirements, docstrings, inline comments, and protocol RFCs—to extract the structural hierarchy and intended business logic ($D_c$).

\item \textbf{Contract Inference:} The agent is instructed to systematically deduce the behavioral and safety boundaries for each method. It must explicitly formalize these limits by populating the pre- ($Pre_c$) and post-conditions ($Post_c$).

\item \textbf{Coverage Assessment:} The agent acts as an analytical cross-referencer, evaluating its inferred conditions against any (optionally) provided unit tests and actively discovering additional relevant tests across the codebase. It marks each condition in $Pre_c$ and $Post_c$ as either \textit{Tested (\checkmark{})} or \textit{Untested (\textbf{X})} based on current test assertions.

\item \textbf{Artifact Generation:} For every \textit{Untested} condition, the agent generates a specific, targeted test suggestion to populate $Sug_c$. The final output of this phase is the collected, semi-formal specification document reflecting the complete $\langle D_c, Pre_c, Post_c, Sug_c \rangle$ tuples for all methods, including helper functions and private methods.

\end{enumerate}

\subsection{Human-in-the-Loop Specification Curation}
\label{sec:hitl}

Because the generated specification artifact is human-readable, our framework naturally supports an \textit{optional} Human-in-the-Loop (HITL) step between the specification extraction and test generation phases.
In this step, the developer can refine the inferred
$Pre_c$ and $Post_c$, as well as accept/reject/amend
$Sug_c$ based on their expertise, before any code is generated.

\subsection{Phase II: Oracle-Driven Test Generation}
\label{sec:design:test}

Once the specification artifact is generated and optionally manually finalized, the agent transitions to the synthesis phase.
Given the finalized specification $S$, and the current test suite $T_{base}$, the agent outputs an augmented test suite $T_{gen}$ that implements the test suggestions ($Sug_c$) corresponding to the \textit{Untested} conditions identified in Phase I.

By gating test generation behind the completion of $S$, the agent's context window is primed with explicit boundary conditions. The agent systematically iterates through $Sug_c$ to synthesize, build, and validate concrete unit assertions, producing a final augmented test suite grounded in behavioral contracts.

\section{Empirical Design}
\label{sec:empirical_design}

To understand how this spec-driven framework can improve agentic test suite generation, we perform a comparative analysis using faults found in industrial production environments.
Our study aims to answer the following research questions:

\begin{itemize}
\item \textbf{RQ1 (Quantitative Effectiveness):} To what extent does grounding an agent with a semi-formal specification improve structural coverage and the rate of detecting real-world, historical bugs?

\item \textbf{RQ2 (Qualitative Rigor):} How do tests generated using and not using semi-formal specifications compare in terms of readability, best practices, and
edge-case coverage?

\item \textbf{RQ3 (Specification Accuracy):} How accurately do the agent-generated semi-formal specifications establish testing oracles that capture the violated behavioral contracts?

\item \textbf{RQ4 (Cost-Efficiency):} What is the token consumption overhead incurred by the Spec-Driven Agent compared to a non-spec-driven, baseline agent?
\end{itemize}

\subsection{Dataset Collection}

We curated a dataset of 90 historical bug-fixes (pairs of buggy and fixed code) from Google's Internal Issue Tracking System.
Similar to foundational evaluation frameworks like Defects4J \cite{just2014defects4j}, we evaluate our approach on real faults because real-world production bugs inherently involve complex implicit state dependencies and boundary violations that stress an AI's code reasoning capabilities far beyond standard generative tasks.

The dataset was strictly filtered to include only human-filed bugs with verified fixes and reproducible runtime failures.
Bugs involving trivial syntactic fixes, multimedia processing, or personally identifiable information were excluded.
Furthermore, we only consider bugs whose fix changes a single production code file, a corresponding test file, and optional \CodeIn{BUILD} configuration files.
The final dataset spans multiple languages, including C++, Java, Python, and Go, providing evidence of language-based generalization
of our findings.

\subsection{Agent Configurations}

To ensure a controlled comparative analysis, we isolate the impact of the specification reasoning phase by deploying both the baseline and spec-driven agents on the exact same underlying agentic architecture. Both agents use the Gemini 3 Flash~\cite{gemini3flash}
and operate within an identical agentic harness equipped with the following tool set: \CodeIn{view\_file}, \CodeIn{list\_dir}, \CodeIn{write\_to\_file}, \CodeIn{run\_command}, \CodeIn{grep\_search}, \CodeIn{find\_by\_name}, and a specialized \CodeIn{run\_test} execution tool.
Both agents use the default Gemini 3 Flash inference parameters: Temperature 1.0, TopP 1.0, and TopK 50.

\paragraph{Baseline Agent}

The baseline is a standard prompt-to-code agent, broadly representative of existing architectures~\cite{cheng2026dynamic,yang2024swe}.
It is provided with the source file under test and explicitly instructed to:

\begin{enumerate}
\item Analyze the code file.
\item Create a comprehensive test suite.
\item Modify the test file and optionally \CodeIn{BUILD} file if necessary.
\item Run the \CodeIn{run\_test} tool with a provided test target to validate its generated implementation.
\end{enumerate}

\paragraph{Spec-Driven Agent}

Our proposed agent uses the same model and tool set but is bound by the framework described in \S\ref{sec:design} -- Instead of direct code synthesis, the agent is forced to construct and validate the intermediate behavioral specification ($S$) before generating tests.

While the original framework supports HITL (\S\ref{sec:hitl}), evaluating the efficacy of human intervention introduces significant subjectivity and confounding variables.
Therefore, we bypass the HITL phase and accept all agent-generated test suggestions autonomously to facilitate a scalable, objective assessment of the model's baseline reasoning capabilities.

\subsection{Greenfield Test Generation Setup}

To evaluate whether the semi-formal specification provide a ``cognitive scaffold'' for the agent to generate tests, we employ a \textit{Greenfield Test Generation} setup, where (by construction) no prior tests are available for the to-be-tested code.
This deliberate isolation prevents the agent from taking heuristic shortcuts---such as copying established test structures or performing simple test repair--and instead forces it to rely entirely on its fundamental code understanding and inferred specification to navigate the complex state space.
In addition, completely removing prior test suites ensures a standardized evaluation baseline across projects of varying initial test quality.

For each bug $b$ in our dataset, let $C_b$ denote the source file in its original buggy state, $C_f$ denote the source file in its fixed state, and $T_b$ denote the pre-existing test suite source file(s). Note that
within Google's build system, $T_b$ is easily retrieved by identifying all test targets that explicitly declare $C_b$ as a direct dependency.

The evaluation procedure follows these steps:

\begin{enumerate}
\item \textbf{Initialize Fixed Source:} The agent's workspace is synchronized to the state immediately following the human fix, establishing the clean implementation $C_f$ from which expected component behavior is extracted. To prevent evaluation tautology (i.e., the agent reverse-engineering tests directly from the bug or fix diff), the agent is kept entirely blind to the bug, the commit message, and the fix diff during both specification and test generation phases.

\item \textbf{Strip Existing Tests (Greenfield):} To enforce a greenfield test generation scenario, we remove the existing test suite $T_b$, resulting in an empty test target $T_\emptyset = \emptyset$.

\item \textbf{Agent Invocation:} Let $A \in \{A_{base}, A_{spec}\}$ represent the agent under evaluation. The agent is provided with the fixed source code $C_f$ and the empty test target $T_\emptyset$ to generate a new test suite $T_{gen} = A( C_f, T_\emptyset )$
\end{enumerate}

\begin{enumerate}
\item \textbf{Validation and Coverage Assessment:} The test suite $T_{gen}$ is executed against the fixed code $C_f$. In this step, we compute the structural coverage, denoted as $Cov_{\text{line}}(T_{gen}, C_f)$
and $Cov_{\text{branch}}(T_{gen}, C_f)$, and record whether the test suite passes on the correct logic (i.e., $Pass(T_{gen}, C_f) = \text{True}$).

\item \textbf{Fail-to-Pass Verification (Bug Detection):} The agent's workspace is reverse-patched to the original buggy implementation $C_b$ (except for the test suite and optionally modified \CodeIn{BUILD} files). The test suite $T_{gen}$ is then executed against $C_b$. We formalize successful bug detection, $D(b, T_{gen})$, as true if and only if at least one individual test $t \in T_{gen}$ fails when evaluated on the buggy code:

\[
D(b, T_{gen}) = \begin{cases}
1, & \text{if } \exists t \in T_{gen} \text{ such that } t(C_b) \rightarrow \text{Fail} \\
0, & \text{otherwise}
\end{cases}
\]
\end{enumerate}

Note that build failures are explicitly not considered as test failures and thus do not contribute to successful bug detection.
The bug detection metric verifies that the generated test suite acts as a valid oracle,  where the agent can recover the contracts and tests that successfully isolate and trigger the historical fault.
The reverse-patching step mirrors the fail-to-pass protocols established in prior literature evaluating both classical \cite{shamshiri2015do, almasi2017industrial} and LLM-based \cite{yin2024what} test generation tools against real-world defects.

\subsubsection{Post-Execution Sanity Checks}

To preserve the integrity of our evaluation, we employ post-execution sanity checks on the agents' file operations.
After each run, we audit the workspace trace and automatically discard (and do not repeat)
any run where the agent modified the source code under test ($C_f$) or any other pre-existing source files within the repository.
This validation step is critical: permitting the agent to alter the source code could inadvertently inject new faults into the fixed implementation, leading to test failures that are artifacts of the agent's tampering rather than authentic detections of the historical bug (i.e., false positives).
For a run to be considered valid, modifications to pre-existing files must be strictly limited to the target test file and, if necessary for compilation, the project's build configuration files.

Nonetheless, to fully accommodate agentic reasoning, planning, and our proposed Phase I contract extraction, both the baseline and spec-driven agents are permitted to freely create and iterate on temporary auxiliary files during their run, such as markdown specifications (\CodeIn{.spec.md}s), task lists, and scratchpads.

\subsection{Evaluation Protocol and Metrics}

\subsubsection{RQ1: Quantitative Effectiveness}
\label{sec:eval_rq1}

To account for the inherent non-determinism of LLM, we perform \textit{k} = 5 independent sampling runs for each bug configuration.
We evaluate the generated test suites across three dimensions to ensure basic validity, breadth of coverage, and depth of bug detection:

\textbf{Test Suite Pass Rate (pass@\textit{k}):} measures the rate at which the entire generated test suite compiles and passes on the \textit{fixed code}.
Evaluated as pass@5, this metric serves as a sanity check before we assess the suite's coverage or fault-finding depth.

\textbf{Fault Detection Rate (detect@\textit{k}):} measures the rate of successfully detecting historical bugs via Fail-to-Pass validation (i.e., whether \textit{at least one} of the \textit{k} = 5 generated test suites passes on the fixed implementation but fails on the reverse-patched buggy state).
While software testing literature frequently relies on mutation scores, synthetic mutants often fail to represent the semantic complexity of real-world developer errors \cite{shamshiri2015do}.
In this study, we conceptualize the historical bug as the ``most valuable mutant,'' making detect@\textit{k} our primary measure of test suite efficacy.

\textbf{Structural Coverage:} We check raw line and branch coverage.

\paragraph{Statistical Tests}
We use the statistical estimation and hypothesis testing procedures below to assess the robustness of our results:

\textbf{Estimation via Bootstrapping:} For pass@\textit{k} and detect@\textit{k} metrics, we perform bootstrapping at the bug level (resampling the 90 unique bugs with replacement) with \textit{n} = 10,000 resamples.
This allows us to calculate the mean estimation alongside the 95\% Confidence Interval (CI), providing a robust measure of effect size and result stability.
This statistical estimation methodology is performed similarly to the task-level bootstrapping utilized in the evaluation of the OpenAI Codex model \cite{chen2021evaluating}.

\textbf{McNemar’s Test for Binary Detection:} For comparing fault detection rates between the baseline and spec-driven agents, we construct a 2x2 contingency table of Baseline Success and Failure versus Spec-Driven Agent Success and Failure.
We apply McNemar’s test to determine if the differences in detection rates are statistically significant with \textit{p} < 0.05.

\textbf{Wilcoxon Signed-Rank Test for Continuous Coverage:}
We use this paired difference test to compare line coverage distributions and branch coverage distributions, given that coverage metrics on paired agent runs are typically non-normally distributed.

\subsubsection{RQ2: Qualitative Test Rigor via LLM-as-a-Judge}
\label{sec:eval_rq2}

We employ an LLM-as-a-Judge to assess different qualitative dimensions of the generated test suites that structural metrics cannot capture.

To mitigate model capability risks from LLM-based assessment, we employ a stronger model (Gemini 3.1 Pro~\cite{gemini31pro}) to judge the outputs of our agents that use Gemini 3 Flash~\cite{gemini3flash}.
To mitigate judging bias, we randomize the order of the two provided test suites and anonymize their origins by denoting them as ``Test A'' and ``Test B'' to the judge.

The judge considers four core qualitative criteria and outputs a detailed rationale and preference (``A'', ``B'', or ``Tie'') for each of the following dimensions:

\textbf{Adherence to Testing Best Practices:} To establish an objective standard, we instructed an LLM to condense Google's internal developer documentation on unit testing best practices into a core set of principles, which were then provided to the LLM judge as an evaluation rubric.
This condensed list emphasizes concepts such as behavior-driven testing (focusing on testing distinct behaviors and states rather than methods), maintaining logic-free assertions
(prioritizing DAMP (``Descriptive and Meaningful Phrases'') over DRY (``Don't Repeat Yourself'') to ensure tests are trivially correct upon inspection~\cite{winters2020software}), and utilizing proper factory helpers for complex object construction.

\textbf{Readability and Structural Clarity:} Descriptive naming conventions and adherence to the Arrange-Act-Assert pattern~\cite{beck2003test}.

\textbf{Edge-Case Coverage:} Strategic targeting of boundary values and exceptional behaviors explicitly identified in the specification.

\textbf{Overall Superiority:} A holistic assessment determining which test suite is superior overall.

We perform dimension-wise majority voting over 5 LLM judge runs per agent run.
If there is a tie or no strict majority is reached among the invocations, the final aggregated result for that bug and category defaults to a ``Tie''.
Finally, to quantify the variance and reliability of the LLM judge, we compute and report the agreement rate across the 5 invocations for each evaluation criteria.

\paragraph{Comparison to Original, Developer-Written Test Suites}

To understand how our approach compares to expert engineers, we also perform the same LLM-as-a-Judge assessment to compare between the Spec-Driven Agent's generated test suites and the original test suites authored by the human developers.

\subsubsection{RQ3: Specification Accuracy and Contract Coverage}

To answer RQ3, we introduce ContractCoverage@\textit{k}. While measuring absolute semantic completeness across all execution paths is practically unfeasible, Contract Coverage serves as a pragmatic, defect-driven proxy evaluating whether the generated contract captures the pre- or post-condition violated by a historical bug. This removes test generation capability as a confounding variable when assessing oracle quality. We note that this metric carries two specific limitations: it requires a ground-truth fix and reproducible bug, and it assesses contract boundaries only relevant to the target defect rather than the entire component's behavioral spectrum.

An
LLM-as-a-judge
evaluates the specifications. Similar to the procedure in \S\ref{sec:eval_rq2}, the judge is invoked 5 times per evaluation, and the final decision regarding Contract Coverage is determined via majority vote. To ensure a comprehensive, high-fidelity evaluation of contract alignment, the judge is provided with: (1) the bug description, (2) the developer's commit message (CL description) capturing the human developer's intent, (3) the unified, ground-truth human fix diff of the source files, and (4) the generated specification. The judge performs the following tasks:

\textbf{Contract Coverage Assessment:} Determining if the generated specification explicitly documents the behavioral constraint that was violated by the historical bug and addressed by the fix.

\textbf{Taxonomy Classification:} Categorizing the nature of the documented contract (e.g., distinguishing between low-level
safety specifications like memory allocation limits, and high-level behavioral specifications like protocol state compliance).

\textbf{Failure Diagnosis:} For bugs that were \textit{not} detected, classifying the root cause either as a ``Specification Generation Failure'', where the testing oracle was missing the contract entirely, or a ``Test Generation Failure'', where the contract was identified but a compiling, fault-revealing test was not generated.

\section{Evaluation Results}
\label{sec:results}

In this section, we present results from our empirical study.

\subsection{RQ1: Quantitative Effectiveness}
\label{sec:results:rq1}

We evaluate whether grounding an agent in behavioral specifications improves its capability to detect real-world bugs and achieve higher structural coverage.
We analyze these aspects using two primary dimensions: fault detection capability (detect@\textit{k}) and structural coverage (line and branch coverage) across multiple independent runs ($k \in [1, 5]$).

\subsubsection{Test Suite Pass Rate (pass@\textit{k})}

We first measure pass@\textit{k} as a sanity check, which measures the rate at which the agent authors syntactically valid and contextually correct test suites that compile and pass entirely on the fixed version of the source code.
This metric acts as a necessary baseline to ensure both agents are capable of producing valid code before evaluating their bug detection capabilities. Table~\ref{tab:test_suite_passonfixed_rate_textitpassk_for_k_in_1_5} reports the pass rates for both baseline and spec-driven agents as \textit{k} increases from 1 to 5.

\begin{table}[h]
\centering
\caption{Test suite pass@\textit{k}. Both agents achieve similarly high rates, validating that spec-driven reasoning does not degrade code generation performance.}
\label{tab:test_suite_passonfixed_rate_textitpassk_for_k_in_1_5}
\resizebox{0.85\columnwidth}{!}{
\begin{tabular}{cccc}
\toprule
k & Baseline Agent & Spec-Driven Agent & Difference \\
\midrule
1 & \textbf{94.4\%} & 94.2\% & -0.2\% \\
2 & \textbf{98.0\%} & 97.8\% & -0.2\% \\
3 & \textbf{98.4\%} & 98.3\% & -0.1\% \\
4 & \textbf{98.7\%} & \textbf{98.7\%} & 0.0\% \\
5 & \textbf{98.9\%} & \textbf{98.9\%} & 0.0\% \\
\bottomrule
\end{tabular}
}
\end{table}

As shown in Table~\ref{tab:test_suite_passonfixed_rate_textitpassk_for_k_in_1_5}, both configurations can generate working test suites. At \textit{k} = 1, both agents achieve a pass rate of over 94\%, which rapidly scales to 98.9\% at \textit{k} = 5. The negligible difference between the approaches (-0.2\% at \textit{k} = 1 and 0.0\% at \textit{k} = 5) confirms
both agents possess an equally robust baseline for code synthesis and as a result
improvements observed in branch coverage and bug detection are driven by the quality of the semantic contract guidance rather than code generation proficiency.

\subsubsection{Fault Detection Rate (detect@\textit{k})}
\label{sec:rq1:faultdetect}

The detect@\textit{k} metric estimates the probability of successfully detecting a historical bug within \textit{k} independent test generation runs.
A bug is considered detected if the generated test suite compiles and passes on the fixed version of the code but triggers test execution failures on the buggy version. Table~\ref{tab:bug_detection_probability_textitdetectk_with_95_confidence_intervals_and_mcnemars_test_textitpvalues_for_k_in_1_5} summarizes the bug detection rates for both agents as \textit{k} increases from 1 to 5, along with their bootstrapped 95\% confidence intervals (CI) and McNemar's test \textit{p}-values.

\begin{table*}[t!]
\centering
\caption{Fault detection rate (detect@\textit{k}) with 95\% confidence intervals and McNemar's test \textit{p}-values.}
\label{tab:bug_detection_probability_textitdetectk_with_95_confidence_intervals_and_mcnemars_test_textitpvalues_for_k_in_1_5}
\resizebox{0.7\textwidth}{!}{
\begin{tabular}{lcccc}
\toprule
k & Baseline Agent & Spec-Driven Agent & Difference & McNemar's p-value \\
\midrule
1 & 36.9\% [28.4\%, 45.3\%] & \textbf{41.1\%} [32.9\%, 49.3\%] & +4.2\% & p=0.3075 \\
2 & 45.3\% [36.2\%, 54.6\%] & \textbf{51.3\%} [42.2\%, 60.3\%] & +6.0\% & p=0.4545 \\
3 & 49.4\% [39.6\%, 59.2\%] & \textbf{56.6\%} [47.0\%, 65.9\%] & +7.2\% & p=0.0574 (Marginal) \\
4 & 51.8\% [41.8\%, 61.8\%] & \textbf{60.5\%} [50.7\%, 70.0\%] & +8.7\% & p=0.0352 (\textbf{Significant}) \\
5 & 53.4\% [43.3\%, 63.3\%] & \textbf{63.2\%} [53.3\%, 73.3\%] & +9.8\% & p=0.0352 (\textbf{Significant}) \\
\bottomrule
\end{tabular}
}
\end{table*}

As shown in Table~\ref{tab:bug_detection_probability_textitdetectk_with_95_confidence_intervals_and_mcnemars_test_textitpvalues_for_k_in_1_5}, the Spec-Driven Agent consistently outperforms the
Baseline Agent at every value of \textit{k}.
At \textit{k} = 1, the Spec-Driven Agent achieves a 41.1\% detection rate, representing a 4.2\% absolute improvement over the baseline (36.9\%).
Notably, the performance gap widens as the number of runs increases.
At \textit{k} = 5, the Spec-Driven Agent reaches a detection rate of 63.2\%, yielding a 9.8\% absolute improvement over the baseline agent (53.4\%).
This widening gap suggests that the spec-driven framework leverages additional execution budgets to guide the agent towards exploring diverse path spaces and uncovering subtle bugs.

\paragraph{Progress on More Challenging Bugs}
It is also worth noting the relatively wide 95\% confidence intervals observed for both approaches in Table~\ref{tab:bug_detection_probability_textitdetectk_with_95_confidence_intervals_and_mcnemars_test_textitpvalues_for_k_in_1_5}.
This broad spread is primarily attributable to the high variance in bug detection difficulty within our dataset.
Rather than a uniform probability of detection across all faults, the per-bug detection rates exhibit a heavily bi-modal distribution: individual historical bugs tend to be either consistently caught across almost all independent runs (low difficulty) or almost never caught in any run (high difficulty).
Specifically, in Baseline Agent, 65.6\% of the bugs (59 out of 90) fall into these two extreme categories of 0\% or 100\% detection rates.
The Spec-Driven Agent reduces this bi-modality slightly to 57.8\% (52 out of 90) by shifting several previously undetected (0\%) bugs into the intermediate region (20\% to 80\% detection rates) where success is conditional.
This shift indicates that the Spec-Driven Agent makes progress on more challenging bugs, converting them from completely undetected ($0\%$ detection rate) to non-deterministically detected ($20\%$ to $80\%$ detection rate). This intrinsic polarization in problem complexity naturally inflates the variance of the overall mean estimation across the sample, resulting in the wider confidence bounds.

\paragraph{Uniquely Detected Bugs}
To understand the uniqueness of the bugs detected by each agent, we analyzed the overlap of successful detections at \textit{k} = 5.
Both agents successfully detected a shared subset of 45 bugs.
The Spec-Driven Agent uniquely detected 12 bugs that the Baseline Agent missed, while the Baseline Agent uniquely detected 3 bugs that the Spec-Driven Agent missed.

\paragraph{Statistical Significance}
We applied McNemar's test to evaluate the statistical significance of the difference in bug detection rates at each \textit{k}, as shown in Table~\ref{tab:bug_detection_probability_textitdetectk_with_95_confidence_intervals_and_mcnemars_test_textitpvalues_for_k_in_1_5}.
At lower execution budgets (\textit{k} = 1 and \textit{k} = 2), the differences are not statistically significant (\textit{p} > 0.05). However, as the execution budget increases, the differences become statistically meaningful.
At \textit{k} = 3, we observe marginal significance (\textit{p} = 0.0574).
At higher execution budgets of \textit{k} = 4 and \textit{k} = 5, the Spec-Driven Agent's improvement over the Baseline Agent becomes statistically significant, yielding a \textit{p}-value of \textit{0.0352} (\textit{p} < 0.05) in both cases.
This demonstrates that grounding the agent's reasoning in explicit behavioral contracts yields superior (statistically significant at higher execution budgets) fault-revealing capability compared to direct prompt-to-code synthesis.

\subsubsection{Structural Coverage}

We evaluate structural coverage using the average line and branch coverage of the target source file achieved by the generated test suites on the fixed code version.
For each target source file, the coverage percentage per run is defined as the number of executed lines (or branches) over the number of instrumented (executable) lines (or branches) in that file.
For each bug, we compute the mean of this coverage across all $k=5$ runs.
Table~\ref{tab:average_line_and_branch_coverage_with_95_confidence_intervals} presents the overall mean coverage results across all bugs, along with their 95\% confidence intervals (estimated via bug-level bootstrapping) and Wilcoxon signed-rank test.

\begin{table*}[ht!]
\centering
\caption{Average line and branch coverage with 95\% confidence intervals.}
\label{tab:average_line_and_branch_coverage_with_95_confidence_intervals}
\resizebox{0.8\textwidth}{!}{
\begin{tabular}{ccccc}
\toprule
Metric & Baseline Agent & Spec-Driven Agent & Difference & Wilcoxon p-value \\
\midrule
Line Coverage & \textbf{74.8\%} [69.6\%, 79.8\%] & 74.4\% [68.6\%, 79.9\%] & -0.4\% & p=0.3659 (N.S.) \\
Branch Coverage & 46.4\% [39.6\%, 53.3\%] & \textbf{48.9\%} [41.4\%, 56.4\%] & +2.5\% & p=0.0034 (\textbf{Significant}) \\
\bottomrule
\end{tabular}
}
\end{table*}

The Spec-Driven Agent achieves slightly lower average line coverage (74.4\% vs 74.8\%, -0.4\%) but notably higher average branch coverage (48.9\% vs 46.4\%, +2.5\% improvement). To assess statistical significance, we performed a Wilcoxon signed-rank test on the paired coverage distributions, as shown in Table~\ref{tab:average_line_and_branch_coverage_with_95_confidence_intervals}, and find that
the branch coverage difference is highly statistically significant (\textit{p}-value=\textit{0.0034}), while
the difference in line coverage is not statistically
significant (\textit{p}-value=\textit{0.3659}).

This indicates that the spec-driven framework systematically drives the agent to cover more complex control flow structures (branches), which are crucial for robust validation, rather than simply inflating line coverage by executing straight-line paths.

\subsection{RQ2: Qualitative Test Rigor}
\label{sec:results:rq2}

\begin{figure}[h]
\centering
\includegraphics[width=1\columnwidth]{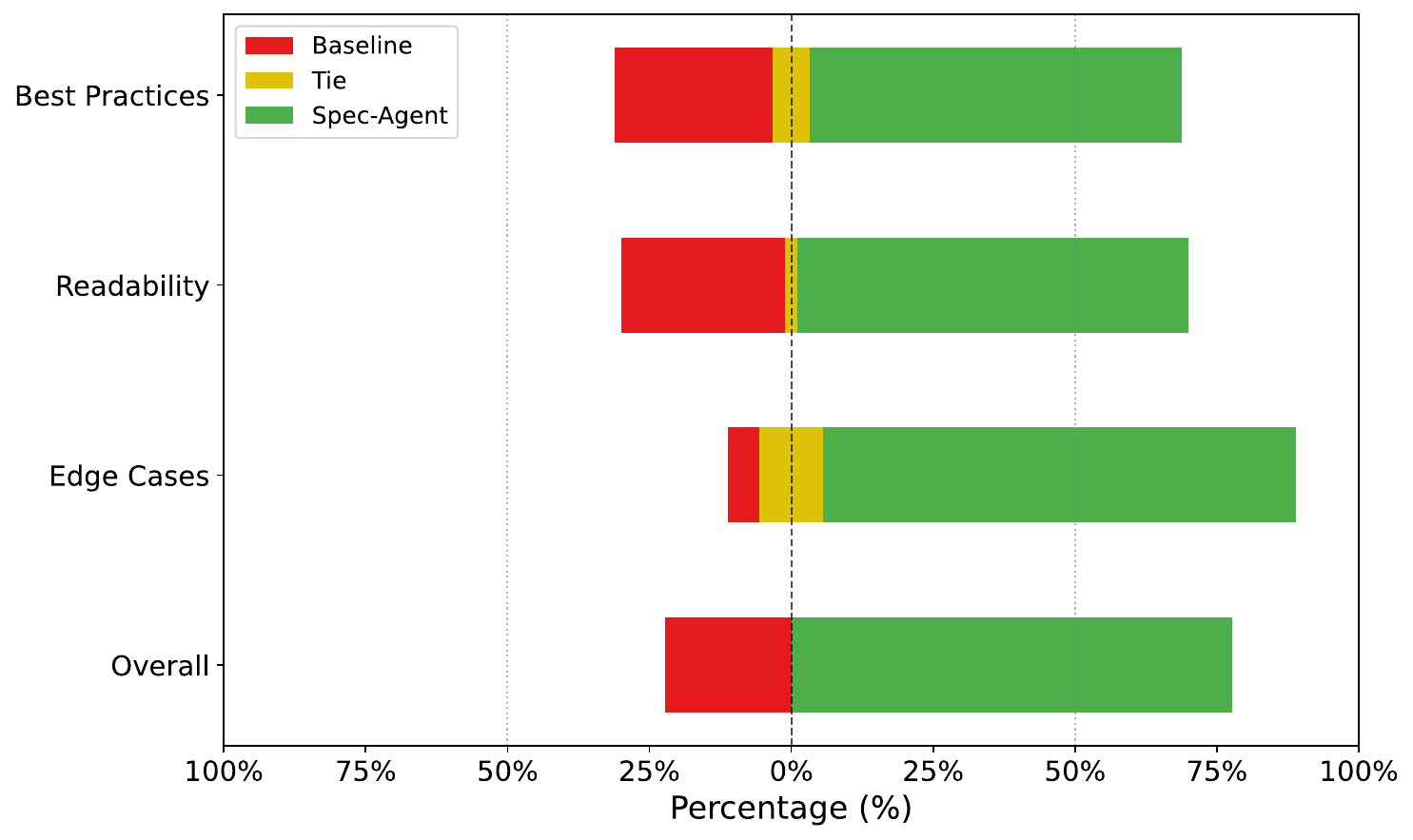}
\caption{Test suites generated by Baseline Agent vs. Spec-Driven Agent. The Spec-Driven Agent produces tests that are consistently judged as higher quality.}
\label{fig:figure_2_rq3_likert_plot}
\end{figure}

\begin{figure}[h]
\centering
\includegraphics[width=1\columnwidth]{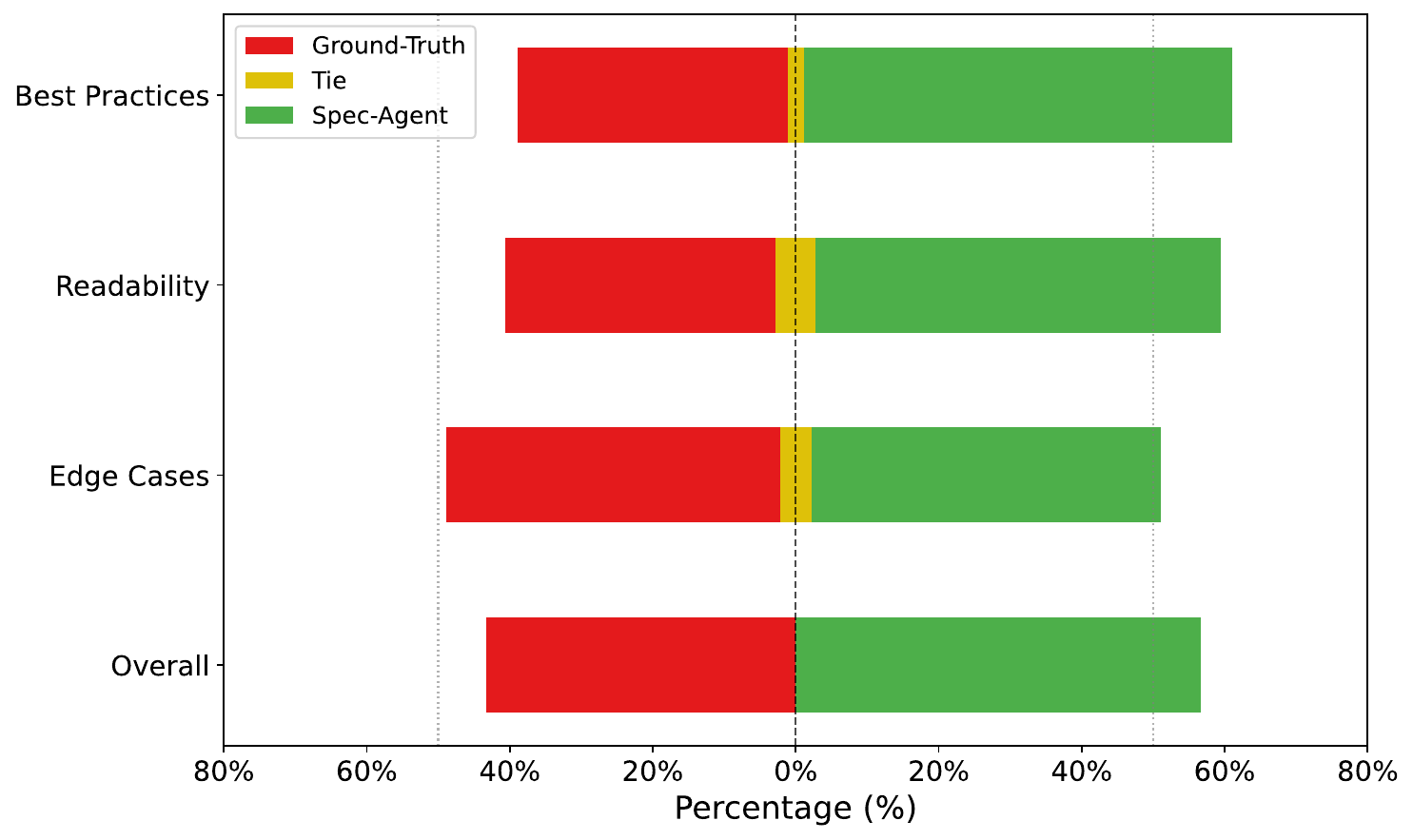}
\caption{Developer-written test suites vs. test suites generated by Spec-Driven Agent. Spec-Driven Agent generates test suites at a quality-level competitive with the original human-written ones, judged via LLM-as-a-Judge.}
\label{fig:figure_3_rq3_likert_plot_ground_truth_developer_tests_vs_specagent}
\end{figure}

We compare the 83 successfully generated
tests from the Spec-Driven Agent against both the test suites generated by the Baseline Agent and the ground truth, developer-written test suites using an
LLM-as-a-Judge.
The judge demonstrated high internal
reliability, achieving high self-agreement ($\ge$90\%)
and an extremely low tie rate ($\le$5.6\%) across 5 independent evaluation runs. Figures \ref{fig:figure_2_rq3_likert_plot} and \ref{fig:figure_3_rq3_likert_plot_ground_truth_developer_tests_vs_specagent} visualize the qualitative evaluation distribution.

\subsubsection{Qualitative Evaluation Results}
We first analyze test suites from the Spec-Driven Agent against those from the Baseline Agent, then against the original test suites written by human developers.

\paragraph{Spec-Driven Agent vs. Baseline Agent}
As shown in Figure~\ref{fig:figure_2_rq3_likert_plot}, test suites from Spec-Driven Agent overwhelmingly outperformed the Baseline Agent across all dimensions on 77.8\% of the cases.
The judge consistently preferred the test suites from Spec-Driven Agent for granular behavioral isolation, idiomatic framework utilization (e.g., avoiding reflection and over-mocking), and contract-aligned naming conventions (e.g., \CodeIn{Method\_State\_Outcome}).
Notably, test suites from Spec-Driven Agent excelled in edge case coverage (83.3\%), systematically validating boundary limits, missing optionality, and error paths that the baseline completely missed.
The higher quality of test suites from Spec-Driven Agent correlated empirically with its 9.8 percentage point advantage in bug detection (\S\ref{sec:rq1:faultdetect}).

\paragraph{Spec-Driven Agent vs. Developer-Written Test Suites} Compared to test suites from human developers  governed by strict style guides and code review, Spec-Driven Agent generated test suites on par or marginally better, being overall superior in 56.7\% of cases, as shown in Figure~\ref{fig:figure_3_rq3_likert_plot_ground_truth_developer_tests_vs_specagent}.
\\

\begin{itemize}
\item \textbf{Agent Strengths:} The Spec-Driven Agent generated narrower, high-precision assertions and minimized boilerplate through localized factory helpers.
Grounded by contracts, it systematically verified negative error paths dictated by the contract.
It also favored robust concurrency constructs (e.g., \CodeIn{absl::Notification}) over the non-deterministic waits occasionally found in human tests.

\item \textbf{Agent Weaknesses:} Human developers excelled at designing exhaustive validation for complex algorithmic scenarios (e.g., intricate state transitions and timezones) that the agent addressed superficially.
Furthermore, humans strictly adhered to public API encapsulation, whereas the Spec-Driven Agent occasionally resorted to reflection-based hacks (e.g., \CodeIn{invokePrivateMethod}) to bypass encapsulation.
\end{itemize}

In summary, across all evaluated qualitative dimensions, the Spec-Driven Agent consistently outperforms the direct generation baseline, while achieving parity with—and occasionally marginally exceeding—the strict rigor of developer-written tests.

\subsection{RQ3: Specification Accuracy}
\label{sec:results:rq3}

We evaluate how accurately the agent-generated semi-formal specifications establish valid testing oracles that capture the violated behavioral contracts. We run the analysis at the attempt level across all $5 \times 90 = 450$ runs.

\subsubsection{Contract Coverage and Bug Detection}
Our analysis of the generated specifications reveals that at a single attempt $(k = 1)$, the Spec-Driven Agent achieves a ContractCoverage@\textit{1} of 61.1\% [53.1\%, 70.7\%], successfully documenting the exact behavioral contract violated in the majority of cases. This coverage rate scales robustly with multiple attempts, rising to 69.7\% [61.3\%, 78.6\%] at $k = 2$, and peaking at a ContractCoverage@\textit{5} of 78.9\% [70.0\%, 86.7\%].

To evaluate the utility of specifications as testing oracles, we constructed a $2 \times 2$ contingency table consisting of binary outcomes of contract coverage and bug detection across all attempts. Because both metrics are binary variables, we evaluate association using the Phi coefficient and assess significance via Fisher's exact test (Fisher's exact test $p = 3.62 \times 10^{-14}$, Phi coefficient $\phi = 0.35$), and we find a statistically significant moderate correlation:

\begin{itemize}
\item \textbf{With Contract Coverage:} When the generated specification successfully documented the violated contract, the agent generated a bug-detecting test suite in 54.9\% of cases (151 out of 275 runs).
\item \textbf{Without Contract Coverage:} When the specification missed the contract, the detection rate dropped to 19.4\% (34 out of 175 runs).
\end{itemize}

This finding provides strong empirical support for our core hypothesis:
\textit{Grounding the agent's reasoning in explicit behavioral contracts significantly improves automated test generation}.
Specifically, when a contract is established in the intermediate specification artifact, the synthesis phase effectively translates it into a fault-revealing test (54.9\%). Conversely, without a contract established, the agent is forced to generate tests without semantic direction, resulting in a lower baseline success rate of 19.4\%.

The remaining 34 runs that detected the bug despite lacking contract coverage represent cases where generic structural coverage goals or simple crash assertions (such as checking for unhandled exceptions) happened to trigger the bug by chance, rather than through targeted oracle-driven validation.

\subsubsection{Specification Generation Failure}

We consider the 141 runs as that failed to detect the bug and where the generated specification did not cover the buggy behavior as \textit{spec generation failures}.
We categorized these failures via an LLM judge into five patterns:

\begin{itemize}
\item \textbf{Omission of Methods (35):} Completely failing to identify and document a specific method, helper, or RPC handler.
\item \textbf{Missing Error Handling (32):} Documenting only the happy path, missing exceptions, fallbacks, and edge cases.
\item \textbf{Omitted Data Transformations (27):} Missing dynamic field mappings, string manipulations, math calculations.
\item \textbf{Abstraction of Constants (25):} Acknowledging constants or formats but abstracting exact values or key mappings.
\item \textbf{Missing Execution Constraints (22):} Missing precise execution ordering, negative constraints, boundary conditions.
\end{itemize}

\subsubsection{Test Generation Failure}

We consider the 110 runs where the contract is deemed covered but the bug is not detected as \textit{test generation failures}.
They are semantic failures where the test suite passed the fixed code but failed to catch the bug, and do not include agent framework or compilation failures (14 runs).
We categorize failures in these runs into five patterns:

\begin{itemize}
\item \textbf{Inadequate Input Data (37):} Providing overly simplistic, trivial, or happy-path inputs without necessary edge-case triggers (e.g., null values, specific timezones).
\item \textbf{Omitted Scenarios (36):} Ignoring explicit specification statements and failing to generate tests for specific branches or methods.
\item \textbf{Weak or Missing Assertions (20):} Executing the buggy path but failing to strictly verify state mutations or using overly broad assertions.
\item \textbf{Flawed Test Logic (12):} Masking bugs with internal test flaws, such as race conditions, over-mocking, or missing trigger actions.
\item \textbf{Invalid Test Code (5):} Generating syntactically incorrect code or targeting non-existent signatures, resulting in compilation failures.
\end{itemize}

\subsection{RQ4: Cost-Efficiency}
\label{sec:tokens}

While grounding agents in explicit behavioral contracts yields substantial improvements in both structural coverage (\S\ref{sec:results:rq1}) and qualitative test rigor (\S\ref{sec:results:rq2}), these benefits must be weighed against the incurred computational cost.
To evaluate this trade-off, we perform a detailed analysis on the token consumption.

Table~\ref{tab:aggregate_token_consumption_and_costefficiency_metrics} presents the aggregate token consumption for both baseline and spec-driven agent configurations across all 5 runs of the evaluated dataset, split by input (prompt) and output (generation) tokens, along with the resulting bug-detection yield.

\begin{table}[ht]
\centering
\caption{Aggregate cost-efficiency metrics.}
\label{tab:aggregate_token_consumption_and_costefficiency_metrics}
\resizebox{\columnwidth}{!}{
\begin{tabular}{cccl}
\toprule
Metric & Baseline & Spec-Driven & Overhead / Ratio \\
\midrule
\textbf{Total Tokens Consumed} & 243.9M & 336.7M & \textbf{1.38$\times$} (+38.0\%) \\
\textit{Input Tokens} & 224.4M & 305.7M & 1.36$\times$ (+36.2\%) \\
\textit{Output Tokens} & 19.5M & 31.0M & 1.59$\times$ (+59.1\%) \\
\textbf{Unique Bugs Detected (\textit{k}=5)} & 48 & 57 & \textbf{+9} (+18.8\%) \\
\textbf{Tokens per Unique Bug Detected} & 5.1M & 5.9M & \textbf{1.16$\times$} (+16.2\%) \\
\bottomrule
\end{tabular}
}
\end{table}

As shown in Table~\ref{tab:aggregate_token_consumption_and_costefficiency_metrics}, the Spec-Driven Agent consumed a total of 336.6M tokens, representing a 38.0\% overhead over the Baseline Agent's 243.9M tokens.
This overhead is a direct consequence of the multi-phase spec-driven test generation framework: the test generation phase (\S\ref{sec:design:test}) carries the overhead of prepending the specification generated from the specification extraction phase (\S\ref{sec:design:spec}) to the agent's context window.

Interestingly, while input tokens increased by 36.2\% (reflecting the larger prompt contexts containing specifications), output tokens exhibited a higher relative increase of 59.1\% (30.9M vs 19.4M).
This increase is driven by the Spec-Driven Agent (1) must output the comprehensive behavioral specification in markdown, and (2) synthesize more robust, thorough, and consequently larger test suites containing extensive edge-case coverage while being guided by the specification.

\section{Threats to Validity}
\label{sec:threats}

\paragraph{Construct Validity}
Our evaluation relies on an LLM-as-a-Judge (Gemini 3.1 Pro) for assessing qualitative test rigor (\S\ref{sec:results:rq2}) and for computing contract coverage (\S\ref{sec:results:rq3}).
A known threat is that LLMs can exhibit classification noise, verbosity bias, and self-preference bias.
We mitigate this by (1) deliberately using a larger, more capable Gemini 3.1 Pro model to evaluate the outputs of a smaller Gemini 3.0 Flash model which helps reduce self-preference bias, and (2) executing the LLM judge 5 independent times for every evaluation pair and computing the final decision via a strict majority vote.
This ensemble approach  smooths classification noise and ensures consistent self-agreement across the independent runs.

\paragraph{Internal Validity}
A pervasive threat when evaluating frontier AI models on historical software engineering datasets is data contamination: the risk that the model encountered the exact evaluation bugs during its pre-training phase, leading to memorization rather than true reasoning.
However, in our evaluation, both the baseline and the Spec-Driven Agent utilize the exact same underlying LLM. Therefore, any observed performance improvements are strictly due to the enhanced reasoning capabilities unlocked by the specification rather than data contamination.

Since our evaluation relies on a single set of LLMs (Gemini), the absolute performance metrics are inherently tied to this version's underlying code-reasoning capabilities, so using different or future LLMs may yield different absolute bug detection and compilation rates. We use the same model set and agentic harness for both evaluated agents to mitigate this threat.

Evaluating bug detection on historical bugs is an established protocol for regression testing. However, we acknowledge that developers want to find new bugs in practical production environments. We focus on understanding the feasiblity of semi-formal specification framework in helping the agent to recover bug-catching contracts, by evaluating on historical bugs.
Given its effectiveness illustrated in this paper, future work can explore using the framework to aid agentic discovery of new bugs.

\paragraph{External Validity}
While our dataset is multi-lingual (C++, Java, Python, Go) and covers diverse domains, all historical bugs were sourced from Google's monorepo, which
represents a single large-scale technology organization with specific engineering standards and style guides. Consequently, our metrics must be interpreted in the context of this specific environment. While the exact detection and coverage rates may vary across different corporate or open-source codebases, the underlying methodology of using semi-formal contracts as a cognitive scaffold for agentic exploration is fundamentally language- and platform-agnostic.
Additionally, our evaluation scope is currently limited to bugs whose fixes modify a single source file to establish cleaner evaluation boundaries and straightforward comparison.

\section{Related Work}
\label{sec:related_work}

We discuss related work in program verification broadly, LLM-based program reasoning, and test generation specifically.

Hoare logic \cite{hoare1969axiomatic} introduced a compositional approach to formally reasoning about program behavior, based on pre- and post-conditions, setting the foundations for modern-day program verification.  The seminal Eiffel programming language \cite{meyer1988eiffel} built on these foundations, introducing design-by-contract, which allows developers to clearly document each component's requirements and responsibilities by co-locating their logical assertions with their implementation and automatically checking these at runtime. Advances in automated reasoning tools and modern verification-aware programming languages, like Dafny \cite{leino2010dafny} and $F^*$\cite{swamy2016dependent}, allow developers to statically verify that their implementations satisfy their logical properties. Similarly, Lean \cite{demoura2015lean} and Rocq \cite{bertot2004interactive} can be used to write rigorous, machine-checkable proofs, allowing computer scientists and mathematicians to prove properties about their programs. Recently, Lean in particular has seen substantial uptake in the AI for math and code research community, with agentic approaches tackling challenges such as autoformalization \cite{zhang2025autoformalization}.

Similar to this line of work, we use pre/post conditions
to specify key program properties. However, our focus is on using these properties to generate robust test suites. We employ an agentic approach to generating the specifications and the test suite, and do not expect users to manually write additional specification properties, nor do we aim to prove full functional correctness (or other important properties, like safety) of the underlying program.

More generally, LLMs have proven to be capable of reasoning about varying types of code properties \cite{yang2025code}. For example, CRUXEval \cite{gu2024cruxeval} introduced a benchmark to evaluate LLM’s reasoning about execution (both forward and backwards). Similarly, RubberDuckBench \cite{mohammad2026rubberduckbench} constructed more complex code reasoning questions. Given the effectiveness of compositional reasoning in software, research has focused on invariant mining and using LLMs to generate preconditions \cite{king2025llm}, post-conditions \cite{endres2024can}, and invariants. Unlike traditional dynamic invariant inference (e.g., Daikon) which captures actual execution traces (including bugs) in rigid mathematical formulas, LLMs leverage contextual signals (docstrings, naming) to capture developer intent in highly readable, semi-formal natural language contracts that generalize across multi-language codebases without runtime setup. For example, HoarePrompt \cite{bouras2025hoareprompt} showed LLMs can produce natural language counterparts to these logical assertions and use them to identify faulty programs that fail to satisfy their original intent. \citet{ugare2026agentic} show that LLM-based agents can build up semi-formal arguments for program equivalence, fault localization, and code question answering. SpecBench \cite{hamblin2026specbench} recently introduced a new benchmark focused on agentic reasoning for program specifications.

In contrast to this work, we are not focused on general program reasoning, such as reasoning about intermediate states or answering code property questions. Our agentic workflow is focused on test generation, and we use specifications as a way to produce more robust tests. While our agent may choose to perform semi-formal reasoning while generating the specifications or the tests, it is not constrained to this particular approach. Finally, we evaluate our approach in the context of a large industrial codebase.

Another line of related work is test generation. Non-learning-based techniques have focused on generating tests that maximize coverage \cite{fraser2011evosuite, pandita2010guided}, while learning-based techniques further consider code naturalness of generated tests \cite{tufano2020unit, watson2020learning, nie2023learning, rao2023catlm}. Recent LLM-based techniques have explored feedback-driven test generation \cite{altmayer2025coverup, dakhel2024effective, mundler2024swt}, change-description guided behavior test generation \cite{pradel2025testora}, bug reproduction test generation \cite{ahmed2025heterogeneous, ahmed2025otter, cheng2026dynamic}, and test suite generation \cite{jain2025testforge}. In contrast to the prior work, we explicitly break down test generation into a specification generation phase and a test generation phase, with the former grounding the latter. Moreover, we focus on evaluating test generation for industrial code, rather than in the context of program repair or new code changes.

Concurrently with our work, \cite{risemsr2026deeptest} introduced DeepTest, which generates tests based on a specification extracted by an agent. While DeepTest builds out its specification by extracting call graphs and analyzing expected relationships (incorporating additional knowledge such as comments and change descriptions), our approach strictly structures the specification as a semi-formal, natural language artifact (\CodeIn{.spec.md}) centered on explicit pre- and post-conditions. Our work is complementary to DeepTest, as we evaluate the performance of this general pattern of grounding test generation on mined specifications in  industrial code.

General-purpose cognitive scaffolding, such as Chain-of-Thought (CoT) reasoning, is already a default behavior in LLMs and is natively utilized by our Baseline Agent. While our work focuses on evaluating semi-formal contracts as a specialized alternative, future work will explore a direct empirical comparison with other structured cognitive scaffolding techniques, such as explicit high-level test planning, pseudocode drafting, or execution-trace modeling. Future work can also complement contract coverage with precision-oriented metrics to identify over-specifications or hallucinations, and compare semi-formal contracts against traditional static analysis and dynamic invariant miners.

\section{Conclusions}
\label{sec:conclusions}

In this paper, we proposed Spec-Driven Test Generation, a novel agentic framework that bridges Design by Contract principles with automated software validation. By instructing agents to explicitly reason about and document a component's behavioral boundaries in a semi-formal specification before generating test code, our approach provides a critical cognitive scaffold that elevates the quality and fault-revealing capability of the generated test suites, while guiding a systematic state-space exploration.

Our empirical evaluation on a dataset of 90 real-world, historical bugs from Google's production systems demonstrates that grounding agents in these self-generated specifications yields statistically significant improvements. At $k=5$ runs, the Spec-Driven Agent achieved a $63.2\%$ bug detection rate ($+9.8\%$ absolute improvement over the baseline) and a significant increase in branch coverage. Furthermore, qualitative evaluations by an LLM-as-a-Judge showed that spec-driven test suites are overall superior to baseline-generated suites in $77.8\%$ of cases, and even outperform or match expert developer-written tests in $56.7\%$ of cases. We also introduced the ContractCoverage@\textit{k} metric to evaluate the semantic completeness of the generated specifications. We observed a significant correlation between achieving this coverage
and downstream bug detection success, highlighting the utility of the specification as a guiding testing oracle.

\balance
\bibliographystyle{ACM-Reference-Format}
\bibliography{main}
\end{document}